\pdfoutput=1

\documentclass[11pt]{article}

\usepackage{acl}
\usepackage{times}
\usepackage{latexsym}
\usepackage{enumitem}
\usepackage{amsmath}
\usepackage{tcolorbox}
\usepackage{tabularx} 
\usepackage{lipsum}   
\usepackage{bm} 

\usepackage[T1]{fontenc}

\usepackage[utf8]{inputenc}

\usepackage{microtype}

\usepackage{inconsolata}

\usepackage{graphicx}
\usepackage{multirow}
\usepackage{booktabs}
\usepackage{array}
\usepackage{float}

\title{Stealing Training Data from Large Language Models \\ in Decentralized Training through Activation Inversion Attack}

\author{Chenxi Dai\thanks{Equal contribution} \and
        Lin Lu\footnotemark[1] \and 
        Pan Zhou\thanks{Corresponding author} \\
  Huazhong University of Science of Technology \\
  \{dcx001,loserlulin,panzhou\}@hust.edu.cn}

\begin{document}
\maketitle
\begin{abstract}
Decentralized training has become a resource-efficient framework to democratize the training of large language models (LLMs). However, the privacy risks associated with this framework, particularly due to the potential inclusion of sensitive data in training datasets, remain unexplored. This paper identifies a novel and realistic attack surface: the privacy leakage from training data in decentralized training, and proposes \textit{activation inversion attack} (AIA) for the first time. AIA first constructs a shadow dataset comprising text labels and corresponding activations using public datasets. Leveraging this dataset, an attack model can be trained to reconstruct the training data from activations in victim decentralized training. We conduct extensive experiments on various LLMs and publicly available datasets to demonstrate the susceptibility of decentralized training to AIA. These findings highlight the urgent need to enhance security measures in decentralized training to mitigate privacy risks in training LLMs.
\end{abstract}

\section{Introduction}

Large language models (LLMs)~\cite{gpt3, chen2023extending, mistral, gemma2} have demonstrated remarkable efficacy across diverse domains~\cite{li2024ecomgpt, wu2024chateda, lu2024chameleon} due to their advanced capabilities in semantic understanding and text generation. However, their emergent abilities follow the scaling law~\cite{bahri2024explaining, naveed2023comprehensive, raiaan2024review}, which leads to state-of-the-art LLMs typically comprising billions of parameters. For instance, the DeepSeek-V3~\cite{liu2024deepseek} model, with its 671 billion parameters, requires 2,664 million H800 GPU hours for training. This resource-intensive training and fine-tuning process presents significant barriers to the democratization of LLMs. As a result, decentralized training~\cite{yuan2022decentralized, ryabinin2023swarm} is gaining increasing attention as a promising solution to mitigate these resource challenges.

Decentralized training is mainly based on parallel training (e.g., \textit{pipeline parallelism}~\cite{narayanan2019pipedream}), which distributes training computations across heterogeneous computing devices (typically GPUs) in a pipeline, with each device acting as a distinct stage. Unlike traditional federated learning (FL), which is based on data parallelism~\cite{li2014scaling, luo2020prague}, pipeline parallelism allocates model layers across devices, facilitating the concurrent processing of multiple data batches over successive stages. During decentralized training, each stage transmits activations during forward propagation and gradients during backward propagation to iteratively update model parameters. This approach enhances memory utilization and alleviates computational bottlenecks. Frameworks such as GPipe~\cite{huang2019gpipe} and Megatron-LM~\cite{narayanan2021efficient} effectively balance resource constraints with training efficiency, supporting the democratization of LLMs.

As research on the robustness of decentralized training progresses, the security vulnerabilities of this framework have become increasingly evident. However, most existing studies~\cite{thorpe2023bamboo, jang2023oobleck, duan2024parcae} primarily focus on addressing fault tolerance issues related to hardware failures in pipeline parallelism, often neglecting the impact of human threats. While some research~\cite{lu2024position} has examined the role of attackers, demonstrating that malicious stages in decentralized training can significantly disrupt training outcomes and hinder model convergence, this study typically assumes that attackers can control any stage of decentralized training. Such strong assumptions about the attackers' capabilities make the attack methods impractical in real-world training scenarios, where tampering with transmitted values is highly likely to be detected by the training initiator. Furthermore, the above studies fail to address privacy risks, which could lead to more severe consequences~\cite{bethany2024large}.

Motivated by this gap, we aim to investigate whether malicious stages in decentralized training can steal privacy without disrupting the training process. However, implementing this privacy reconstruction attack presents a significant challenge: decentralized training differs substantially from traditional training methods, such as localized training or FL. In traditional training, attackers may have access to a complete model copy~\cite{li2023sentence,morris2023text} or its inputs and corresponding outputs~\cite{huang2024transferable}. In contrast, within the decentralized training, malicious stages can only access the transmitted values between stages. This raises a critical research question: \textit{How to steal privacy, such as training data, solely through transmitted values in decentralized training?}

To address this critical research question, this paper first introduces the \textbf{\textit{\underline{A}ctivation \underline{I}nversion \underline{A}ttack}} (AIA) targeting decentralized training. Specifically, we demonstrate how a malicious stage in decentralized training can steal training data by exploiting activations through a two-step process. In the first step: \textbf{Shadow Dataset Construction}, the attacker creates a shadow dataset of text-activation pairs using a public dataset, aiming to align the data distribution of the shadow dataset with that of the actual training process. In the second step: \textbf{Attack Model Training}, the attacker trains a generative model using the shadow dataset to learn the mapping from activations to text labels. The attacker then reconstructs the corresponding training data from victim activations. In summary, the contributions of this paper are as follows:

\begin{itemize}[nolistsep, leftmargin=*, topsep=0pt]

    \item We identify a novel attack surface, marking the first attempt to steal private training data within decentralized training frameworks.

    \item We propose a two-step attack framework, AIA, that steals training data through activations in decentralized training without detection.

    \item We conduct a comprehensive evaluation of the effectiveness of AIA, demonstrating its character-level capability for training data reconstruction. Specifically, AIA achieves 62\% accuracy in stealing private emails when fine-tuning GPT2-XL.
    
\end{itemize}

\section{Related Work}

\subsection{Decentralized Training Safety}

\citet{yuan2022decentralized} initially explores decentralized training for LLMs. Several studies then examine decentralized training in slow networks~\cite{ryabinin2023swarm, wang2023cocktailsgd} and explore the development of geo-distributed training systems tailored for LLMs~\cite{gandhi2024improving, tang2024fusionllm}. While safety concerns in decentralized training have been identified in previous works~\cite{tang2023fusionai, borzunov2022training}, most existing research focuses mainly on ensuring seamless pipeline operations on preemptible devices, employing techniques such as model backup and redundant computation~\cite{thorpe2023bamboo, jang2023oobleck}. \citet{lu2024position} comprehensively evaluate the potential threats in decentralized training. However, the proposed \textit{forward attack} can be easily mitigated by detection methods, making it impractical in real-world scenarios.


\subsection{Data Leakage from Transmitted Values}

\noindent{\textbf{Data leakage from gradients.}}
In the context of FL, researchers such as \citet{zhu2019deep} have explored deep gradient leakage attacks on both visual and language models. 
\citet{balunovic2022lamp} uses auxiliary language models to model prior probabilities, reducing the loss through alternating continuous and discrete optimization. \citet{gupta2022recovering} first recovers a set of words from gradients, and then reconstructs the sentence from this set of words using beam search. \citet{fowl2022decepticons} and \citet{boenisch2023curious} propose a powerful threat model in which the server is malicious and can manipulate model weights, easily reconstructing the data.
\citet{wu2023learning} proposes a simple adaptive attack method that can bypass various defense mechanisms, including differential privacy and gradient compression, and successfully reconstruct the original text.


\noindent{\textbf{Data leakage from embeddings.}}
Another line of research focuses on embedding inversion attacks, where the attacker aims to reconstruct text from embedding representations. \citet{song2020information} reconstructs 50\%-70\% of the input words from embedding models. However, word-level information alone is insufficient to fully reconstruct privacy. \citet{li2023sentence} proposes a generative embedding inversion attack that reconstructs sentences similar to the original input from embeddings. \citet{morris2023text} utilizes an iterative correction approach to reconstruct text information. \citet{huang2024transferable} investigates a black-box attack scenario, reducing the discrepancy between the surrogate model and the victim model through adversarial training. These studies assume that the victim model is fully trained and static, allowing the attacker to access the input sentence embeddings from the victim model, build a shadow dataset, and then train an attack model to reconstruct the original text. However, in decentralized training settings, the malicious stage only has access to a portion of the model, and thus cannot directly access the victim model.

\section{Preliminaries}

\subsection{Threat Model}
\label{sec:threat_model}

\noindent{\textbf{Attack scenario.}}
We consider a decentralized training scenario where the user intends to fine-tune a pre-trained model ${M}_\text{pre}$ using their private dataset $\mathcal{D}_\text{vic}$, resulting in a fine-tuned model ${M}_\text{fine}$. The framework consists of $K$ stages, where $M_i$ represents the sub-layers (e.g., decode layers in LLMs) of the $i$-th stage. During training iteration $t$, $M_i$ transmits activations $\bm a_i^{(t)}$ to $M_{i+1}$ and gradients $\bm g_i^{(t)}$ to $M_{i-1}$. However, an unmonitored decentralized training framework may introduce an honest-but-curious stage as an attacker.

\noindent{\textbf{Attacker's goals.}}
The attacker's objective is to reconstruct character-level training data $\bm d^{(t)}$ from $\mathcal{D}_\text{vic}$ during iteration $t$ in victim decentralized training. Additionally, the attacker seeks to conceal their malicious activities, executing the attack without disrupting the training process to avoid detection by the training initiator or other detection mechanisms.

\noindent{\textbf{Attacker's knowledge.}}
We assume the attacker, as the $i_\text{att}$-th stage, has access to all information related to its own stage, including the sub-layers $M_{i_\text{att}}$ and transmitted data $\bm a_{i_\text{att}}$ and $\bm g_{i_\text{att}}$. This enables the attacker to infer the architecture of ${M}_\text{fine}$ based on the structure of $M_{i_\text{att}}$. However, the attacker is assumed to have no access to other training-related information, such as transmitted data between benign stages or auxiliary information about the training data. This assumption is realistic, as it facilitates the deployment of this attack in real-world decentralized training environments.

\subsection{Motivation}
\label{sec:act_cos}

In Section \ref{sec:threat_model}, it is established that attackers can only reconstruct training data through the transmitted values during the victim model's training process, such as activations and gradients. This section discusses the challenges of using gradients to conduct such attacks and explores the feasibility of using activations to achieve similar objectives.


In decentralized training, traditional deep gradient leakage attacks encounter a significant limitation: the unavailability of the global model and global gradients. Previous researches~\cite {zhu2019deep, gupta2022recovering, balunovic2022lamp} focus on training or searching for a set of texts that, through the victim model’s gradient, approximate the leaked gradient to reconstruct private data. However, in decentralized training, each stage only has access to a partial model and gradients, making it difficult to reconstruct data through gradients.

\begin{figure}[t]
  \includegraphics[width=\linewidth]{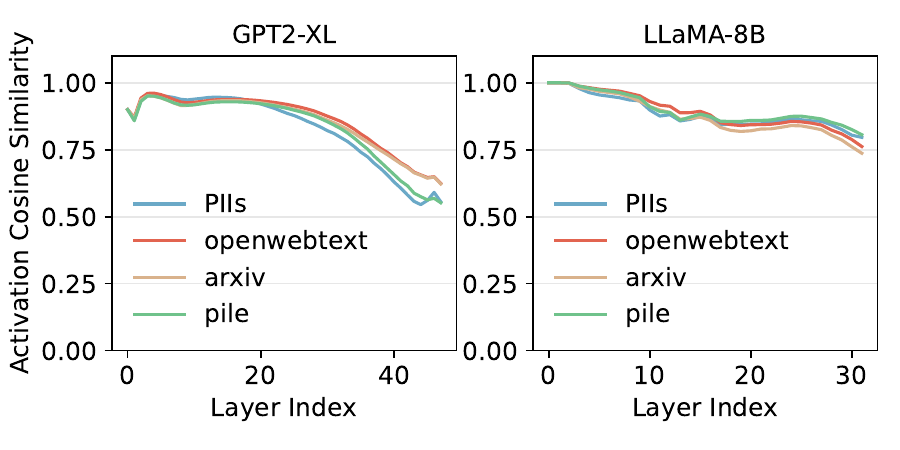} 
  \caption {Cosine similarity between activations for the same data in the pre-trained model and the fine-tuned model across layer index.}
  \label{fig:layer_idx_act_cos}
  \vspace{-1em}
\end{figure}

In contrast, reconstructing data using the intermediate outputs of the victim model is much more straightforward, as these intermediate outputs can be directly used as inputs to train the attack model~\cite{pasquini2021unleashing,li2023sentence}. Inspired by this, we examine the cosine similarity between $\bm a_i^{(t)}$ for $\bm d^{(t)}$ in ${M}_\text{pre}$ and ${M}_\text{fine}$ across layer index $i$ (experimental details can be found in Section \ref{sec:experiment_setup}). As shown in Figure \ref{fig:layer_idx_act_cos}, activation similarity in early layers approaches 100\%, while similarity in later layers remains above 50\%. These results suggest that the activations of the same data exhibit minimal variation before and after fine-tuning, indicating a strong correlation between activations and the training data. This preliminary experiment provides key insights for our attacks in Section \ref{sec:AIA}.

\section{AIA: \textit{Activation Inversion Attack}}
\label{sec:AIA}
\begin{figure*}[t]
  \includegraphics[width=\textwidth]{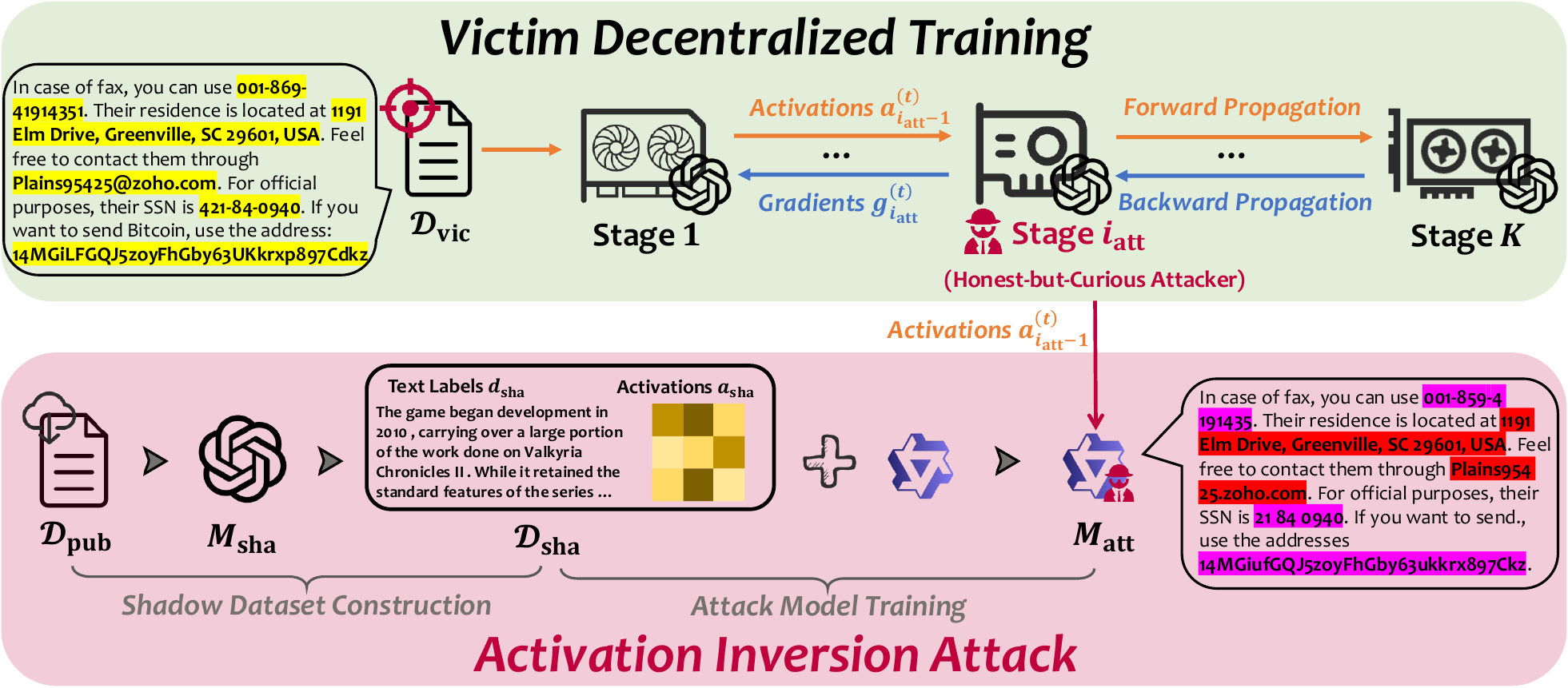}
  \caption{Overview of Activation Inversion Attack (AIA). In a decentralized training system, the victim model $M_{\text{vic}}$ undergoes fine-tuning using private data $\mathcal{D}_{\text{vic}}$, which may contain personally identifiable information values (highlighted in yellow). An honest-but-curious attacker controlling the $i_{\text{att}}$-th stage of the pipeline: (1) records intermediate activation values $\bm a_{i_\text{att}-1}^{(t)}$ captured during the training process, and (2) collects shadow activations $\mathcal{D}_{\text{sha}}$ from the shadow model $M_{\text{sha}}$ to train the attack model $M_{\text{att}}$. Finally, the attacker uses $M_{\text{att}}$ to reconstruct the private data $\mathcal{D}_{\text{vic}}$, with the red and purple text representing precisely recovered and mostly recovered PII data, respectively.}
  \label{fig:system}
  \vspace{-1em}
\end{figure*}

We introduce AIA, a framework for training data reconstruction through activations in decentralized training. During the victim model training, an attacker at the $i_\text{att}$-th stage has access to the activations $\bm a_{i_\text{att}-1}^{(t)}$ passed from $M_{i_\text{att}-1}$ during forward propagation. We denote the mapping function from the original training data $\bm d_\text{vic}^{(t)}$ to $\bm a_{i_\text{att}-1}^{(t)}$ as $f_{[1:i_\text{att}-1]}^{(t)}(\cdot)$. Therefore, we can conclude that: 
$$
\bm a_{i_\text{att}-1}^{(t)}=f_{[1:i_\text{att}-1]}^{(t)}(\bm d_\text{vic}^{(t)})
$$
The attacker's goal can thus be simplified to constructing a mapping function $\phi \approx  (f_{[1:i_\text{att}-1]}^{(t)})^{-1}(\cdot)$ that reconstructs $\bm d_\text{vic}^{(t)}$ from $\bm a_{i_\text{att}-1}^{(t)}$. AIA adopts a learning-based approach by training a generative model to perform this reconstruction. In simple terms, AIA consists of two steps: (1) \textbf{Shadow Dataset Construction}: The attacker first generates a shadow dataset containing text labels and corresponding activations leveraging a public dataset. (2) \textbf{Attack Model Training}: The attacker then uses $\mathcal{D}_\text{sha}$ to train a generative attack model ${M}_\text{att}$ that learns the mapping function $\phi$. Finally, the attacker inputs the actual activations transmitted during the victim model training into ${M}_\text{att}$ to reconstruct the training data. We provide a detailed description of these two steps in the following.

\subsection{Step 1: Shadow Dataset Construction}

Since the attacker cannot access $\mathcal{D}_\text{vic}$, a straightforward approach is to construct a shadow dataset $\mathcal{D}_\text{sha}$ using a public dataset $\mathcal{D}_\text{pub}$. Specifically, we use the frozen pre-trained LLM $M_\text{pre}$ as the shadow model $M_\text{sha}$, with the same type of the victim model, to generate shadow activations $\bm a_\text{sha}$, i.e., 
$$
\bm a_\text{sha}=M_{\text{sha}[1:i_\text{att} -1]}(\bm d_\text{pub})
$$
where $\bm d_\text{pub} \in \mathcal{D}_\text{pub}$. The rationale for this approach is analyzed in Section \ref{sec:act_cos}: the generalizability of $M_\text{pre}$ ensures that the activations remain relatively stable when fine-tuning the victim model $M_\text{vic}$ on $\mathcal{D}_\text{vic}$, allowing us to directly leverage the pre-trained weights from HuggingFace as $M_\text{sha}$. In other words, no additional effort is required to train $M_\text{sha}$, significantly reducing the cost of AIA.

\subsection{Step 2: Attack Model Training}
Next, we focus on training ${M}_\text{att}$ using the shadow dataset $\mathcal{D}_\text{sha}=\{(\bm a_\text{sha}, \bm d_\text{pub})\}$. ${M}_\text{att}$ is designed to take activations as input and output the distribution probabilities of the generated text. It consists of a set of decoder layers and an \texttt{lm\_head} layer. Structurally, it differs from a standard language model by the absence of the initial embedding layer. 
Similar to the recent work~\cite{li2023sentence}, the training objective is to minimize the standard language model loss using teacher forcing~\cite{williams1989learning}:
$$
  L = - \sum_{k=1}^{N} \log P(y_k | x_1, x_2, \dots, x_{k-1})
$$
where $y_k$ is the target word, and $x_i$ represent the input activations. 
Finally, we input the activations $\bm a_{i_\text{att}-1}^{(t)}$ to ${M}_\text{att}$ and obtain $\bm d_\text{vic}^{(t)}$.


\section{Experiments}
\subsection{Experimental Setup}
\noindent\textbf{Victim models.}
\label{sec:experiment_setup}
We conduct experiments on three models: GPT2-XL~\cite{gpt2}, Bloom-7B1~\cite{bloom}, and LLaMA3-8B~\cite{llama3}, which have 48, 30, and 32 decoder layers, respectively. We directly download the pre-trained models from HuggingFace and use them as $M_\text{sha}$ to collect $\mathcal{D}_\text{sha}$. 
To investigate the effects of AIA under extreme conditions, we fine-tune $M_\text{vic}$ for 5 epochs on the corresponding dataset to induce overfitting on the privacy data, thereby maximizing the feature gap between $\mathcal{D}_\text{vic}$ and $\mathcal{D}_\text{sha}$.
The training process is divided into 6 stages, with the assumption that the third stage is malicious. 
The architecture of the attack model is identical to that of the victim model, with all attack models set to 12 decoder layers. 

\begin{figure}[t]
  \includegraphics[width=\columnwidth]{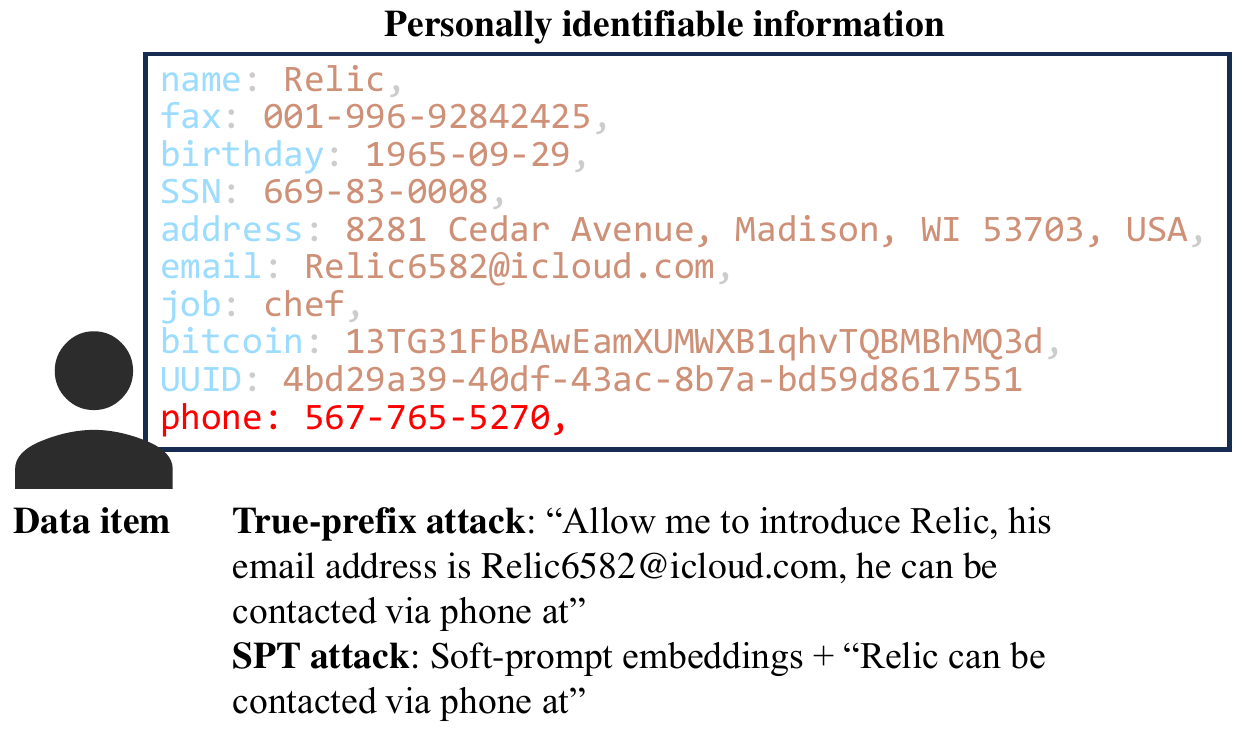}
  \caption{An example of PII data and baseline attacks. The private data includes information such as names, phone numbers, and email addresses. The True-Prefix attack leverages other private attributes to prompt the model to generate the target private attribute, while the SPT attack employs a trained soft prompt added before the query template to extract private information.}
  \label{fig:PII_data_example}
  \vspace{-1em}
\end{figure}

\noindent\textbf{Datasets.}
We use the WikiText~\cite{wikitext} dataset as the attacker's known dataset $\mathcal{D}_\text{pub}$ to construct the shadow dataset $\mathcal{D}_\text{sha}$. The victim datasets $\mathcal{D}_\text{vic}$ include ArXiv, OpenWebText~\cite{openwebtext}, The Pile~\cite{pile}, and a public PII dataset\footnote{https://github.com/zzzzsdaw/PII-dataset}, which contains sensitive information.  An example of a PII data item is shown in Figure \ref{fig:PII_data_example}.

\noindent\textbf{Baselines.}
In the privacy leakage experiments, we adopt the following two methods as baselines. The two methods do not apply to decentralized training, we use them solely for comparison to illustrate the potential risks of our attack. Their attack examples can be seen in Figure \ref{fig:PII_data_example}.

\begin{itemize}[nolistsep, leftmargin=*, topsep=0pt]

    \item \textit{True-Prefix Attack}~\cite{true-prefix} utilizes real prefixes from $\mathcal{D}_\text{vic}$ to prompt the model. In our experiments, we use real PII data of other types within each PII item as the prompt, attempting to induce the model to output the value of the target PII type.

    \item \textit{SPT Attack}~\cite{SPT} trains an additional set of prompt embeddings, which are appended to the original query template. We train the prompt embeddings using 64 PII data pairs, during which the victim model remains frozen and does not require gradient updates.
    
\end{itemize}



\noindent\textbf{Evaluation metrics.}
To evaluate the quality of text reconstruction, we employ the following four metrics.

\begin{itemize}[nolistsep, leftmargin=*, topsep=0pt]

    \item \textit{Perplexity}~\cite{perplexity} assesses the model's capability by measuring the probability distribution of its outputs, with lower values indicating better performance.

    \item \textit{ROUGE}~\cite{rouge} measures the similarity between the generated text and reference text by comparing overlapping words or phrases.

    \item \textit{BLEU}~\cite{bleu} evaluates the similarity between generated text and reference text based on n-gram overlap and is commonly used in machine translation tasks.

    \item \textit{Embedding cosine similarity} calculates the semantic similarity between the generated text and reference text using the all-MiniLM-L6-v2 model\footnote{https://huggingface.co/sentence-transformers/all-MiniLM-L6-v2}~\cite{minilmv2}.
    
\end{itemize}



In the privacy leakage experiments, we evaluate the \textit{attack success rate (ASR)} of our AIA method and two baselines in precisely recovering the values of the target PII types. Precise recovery is defined as correctly outputting the digits and letters in the correct order. During the matching process between the generated data and the original private data, spaces and special characters, such as '-', are ignored, as they do not affect the identification of private data values. The \textit{ASR} is calculated as the ratio of the number of precisely recovered data entries to the total amount of data.

\subsection{Text Reconstruction}

\begin{table*}[ht]
\centering
\caption{Text reconstruction performance of GPT2-XL, Bloom-7B1, and LLaMA3-8B on four datasets. For all metrics except PPL, higher values indicate better performance.}
\label{tab:base_result}
\resizebox{0.94\textwidth}{!}{\begin{tabular}{ccc ccc ccc c}
\toprule[2pt]
\multirow{2}{*}{\textbf{Victim Model}} & \multirow{2}{*}{\textbf{Dataset}} & \multirow{2}{*}{\textbf{PPL}} & \multicolumn{3}{c}{\textbf{ROUGE}} & \multicolumn{3}{c}{\textbf{BLEU}} & \multirow{2}{*}{\textbf{COS}} \\ 
\cmidrule(lr){4-6} \cmidrule(lr){7-9}
                              &                          &                      & \textbf{ROUGE-1}   & \textbf{ROUGE-2}   & \textbf{ROUGE-L}    & \textbf{BLEU-1}    & \textbf{BLEU-2}    & \textbf{BLEU-4}    &                      \\ \hline
\multirow{4}{*}{GPT2-XL}      & PIIs                     & 3.73                 & 0.84     & 0.74     & 0.84      & 0.77     & 0.71     & 0.59     & 0.89                 \\
                              & openwebtext              & 3.09                 & 0.95     & 0.90     & 0.95      & 0.88     & 0.84     & 0.77     & 0.94                 \\
                              & arxiv                    & 5.43                 & 0.92     & 0.85     & 0.92      & 0.81     & 0.75     & 0.64     & 0.92                 \\
                              & pile                     & 1.65                 & 0.98     & 0.95     & 0.98      & 0.95     & 0.93     & 0.89     & 0.97                 \\ \hline
\multirow{4}{*}{Bloom-7B1}    & PIIs                     & 14.82                & 0.80     & 0.67     & 0.80      & 0.67     & 0.60     & 0.47     & 0.89                 \\
                              & openwebtext              & 4.64                 & 0.95     & 0.92     & 0.95      & 0.89     & 0.86     & 0.80     & 0.95                 \\
                              & arxiv                    & 15.45                & 0.91     & 0.83     & 0.90      & 0.77     & 0.70     & 0.56     & 0.90                 \\
                              & pile                     & 2.09                 & 0.97     & 0.95     & 0.97      & 0.95     & 0.93     & 0.90     & 0.95                 \\ \hline
\multirow{4}{*}{LLaMA3-8B}    & PIIs                     & 7.36                 & 0.80     & 0.67     & 0.79      & 0.73     & 0.66     & 0.54     & 0.77                 \\
                              & openwebtext              & 6.50                 & 0.93     & 0.88     & 0.93      & 0.88     & 0.84     & 0.77     & 0.88                 \\
                              & arxiv                    & 9.26                 & 0.88     & 0.78     & 0.88      & 0.80     & 0.73     & 0.60     & 0.83                 \\
                              & pile                     & 2.18                 & 0.96     & 0.93     & 0.96      & 0.94     & 0.92     & 0.89     & 0.92                 \\ \bottomrule[1.5pt]
\end{tabular}}
\vspace{-1em}
\end{table*}

Table~\ref{tab:base_result} presents the performance of AIA across different victim LLMs and datasets. The results indicate that the perplexity of the generated sentences remains below 20, with most values under 10, suggesting that the reconstructed text is relatively fluent and closely aligns with the original fine-tuning data. Both ROUGE-1 and BLEU-1 scores exceed 0.7, with the highest result reaching nearly 0.95, which confirms that the majority of words from the original fine-tuning data are accurately recovered. ROUGE-L scores are generally higher than ROUGE-2, indicating that the generated text maintains high global similarity while exhibiting slightly lower local continuity. However, this slight discontinuity in certain lexical elements has minimal impact on human readability. We further compute the cosine similarity between the embeddings of the generated text and the original text, with values ranging from 0.77 to 0.96, confirming a high level of semantic similarity. These results validate the effectiveness of AIA in reconstructing the original fine-tuning data.

\subsection{Privacy Leakage}
\begin{table*}[ht]
    \centering
    \begin{minipage}{0.285\textwidth}
        \centering
        \makeatletter\def\@captype{table}\makeatother
        \caption{Comparison of the ASR between our AIA method and baselines in stealing phone and email data.}
\label{tab:pii_compare}
        \resizebox{\textwidth}{!}{
        \begin{tabular}{cccc}
\toprule[2pt]
\multirow{2}{*}{\textbf{Victim Model}} & \multirow{2}{*}{\textbf{Method}} & \multicolumn{2}{c}{\textbf{ASR}} \\
                              &                         & \textbf{phone}      & \textbf{email}      \\ \hline
\multirow{3}{*}{GPT2-XL}      & True-Prefix             & 0          & 0.04       \\
                              & SPT                     & 0          & 0.02       \\
                              & AIA(ours)                    & 0.25       & 0.55       \\ \hline
\multirow{3}{*}{Bloom-7B1}    & True-Prefix             & 0.01       & 0.18       \\
                              & SPT                     & 0          & 0.10       \\
                              & AIA(ours)                    & 0.42       & 0.62       \\ \hline
\multirow{3}{*}{LLaMA3-8B}    & True-Prefix             & 0          & 0          \\
                              & SPT                     & 0          & 0          \\
                              & AIA(ours)                    & 0.16       & 0.42       \\ \bottomrule[1.5pt]
\end{tabular}
}
    \end{minipage}%
    \hfill 
    \begin{minipage}{0.665\textwidth}
        \setcounter{table}{4}
        \centering
        \makeatletter\def\@captype{table}\makeatother
        \caption{The impact of attack model architecture on the attack performance of AIA. Each attack model is configured with 6 decoder layers. The results are presented in terms of perplexity.}
\label{tab:attack_model}
        \resizebox{\textwidth}{!}{
        \begin{tabular}{ccc|cccc}
\toprule[2pt]
\multirow{2}{*}{\textbf{Victim Model}} & \multirow{2}{*}{\textbf{\begin{tabular}[c]{@{}c@{}}Attack Model\\ Architecture\end{tabular}}} & \textbf{Shadow Datasets} & \multicolumn{4}{c}{\textbf{Victim Datasets}}                          \\
                                       &                                                                                               & \textbf{wikitext}        & \textbf{PIIs} & \textbf{openwebtext} & \textbf{arxiv} & \textbf{pile} \\ \hline
\multirow{3}{*}{GPT2-XL}      & Mistral                                    & 1.53            & 117.45  & 44.14       & 109.31  & 24.54  \\
                              & Qwen2.5                                    & 1.71            & 410.47  & 115.35      & 301.26  & 68.74  \\
                              & GPT2                                       & 1.54            & 4.17    & 2.61        & 3.81    & 1.70   \\ \hline
\multirow{3}{*}{Bloom7B1}     & Mistral                                    & 1.54            & 7277.80 & 537.97      & 1203.97 & 445.71 \\
                              & Qwen2.5                                    & 1.48            & 7404.53 & 839.47      & 1947.76 & 651.55 \\
                              & Bloom                                      & 1.41            & 16.81   & 9.14        & 13.45   & 2.12   \\ \hline
\multirow{3}{*}{LLaMA3-8B}    & Mistral                                    & 2.60            & 2016.21 & 447.20      & 692.70  & 134.76 \\
                              & Qwen2.5                                    & 2.89            & 1810.44 & 549.28      & 1315.82 & 151.34 \\
                              & LLaMA                                      & 1.85            & 12.57   & 4.16        & 10.11   & 2.03   \\ \bottomrule[1.5pt]

\end{tabular}
}
    \end{minipage}
\end{table*}

\noindent{\textbf{Results compared with baselines.}}
We compare the ASR of AIA with the baselines on the PII types of email and phone, with the detailed results presented in Table~\ref{tab:pii_compare}. The findings indicate that our method performs effectively on both phone numbers and email addresses. For instance, the Bloom-7B1 model achieves precise recovery rates of 41\% for phone numbers and 61\% for email addresses. Even the relatively less effective LLaMA3-8B model accurately recovers 15\% of phone numbers and 41\% of email addresses. 

In contrast, the \textit{True-Prefix Attack} and \textit{SPT Attack} exhibit poor performance, showing minimal success in recovering phone numbers. On the Bloom-7B1 model, both baselines recover only a small portion of email addresses, with ASR of 18\% and 10\%, respectively. We hypothesize that this discrepancy arises from the structure of the PII dataset, where email prefixes consist of a person's name combined with random numbers, enhancing the model's memory of the email. The GPT2-XL model recovers only 2\% to 4\% of email addresses, significantly lower than Bloom-7B1, likely due to its smaller size and weaker capacity for data retention. Notably, neither baseline is able to recover any private data accurately on the LLaMA3-8B model. This may be attributed to the LLaMA3-8B model's alignment and data protection mechanisms implemented during pre-training, which results in the frequent generation of placeholders such as “[email protected]”.


\begin{figure*}[t]
  \includegraphics[width=\textwidth]{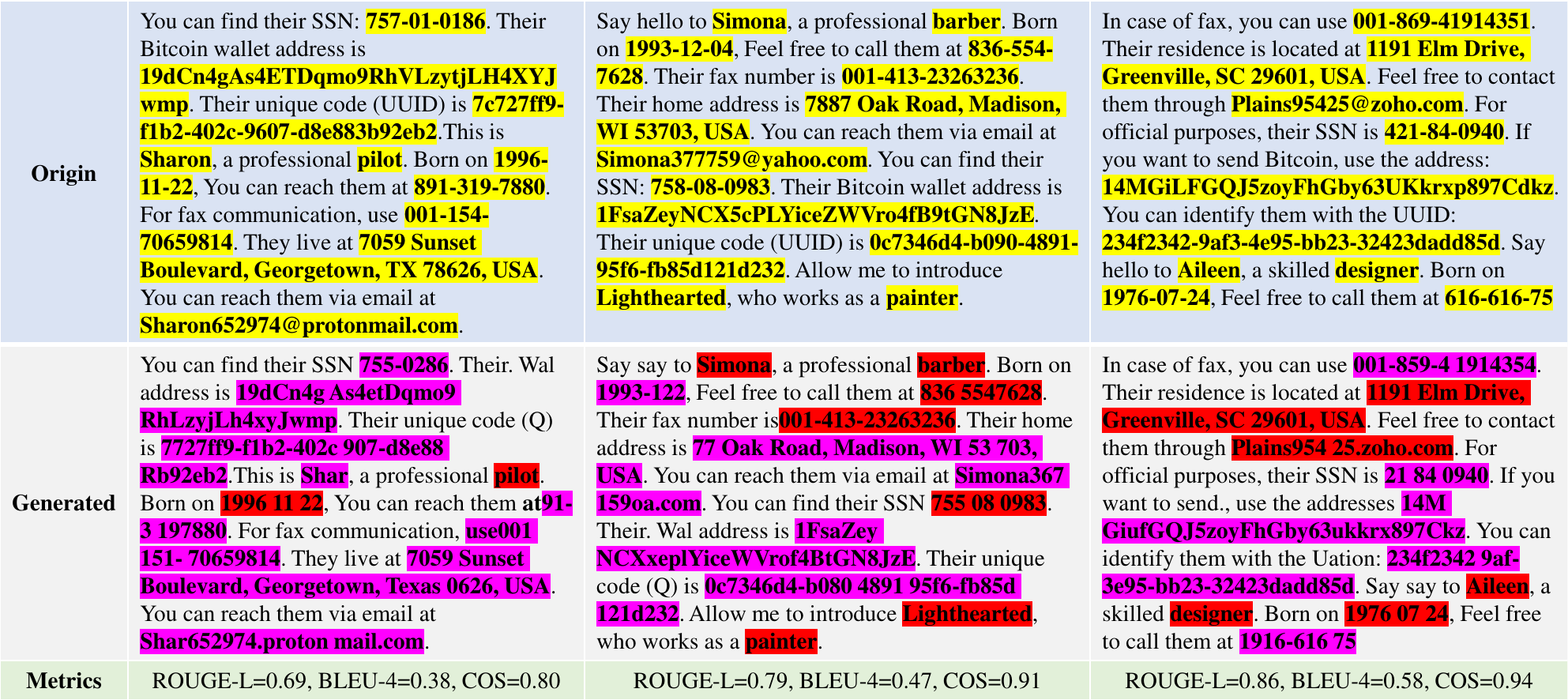}
  \caption{Three comparative examples of generated texts versus original data. The yellow text represents the original PII data, while the red and purple texts represent precisely recovered and mostly recovered PII data, respectively. The text recovery performance improves from left to right.}
  \label{fig:PII_attack_example}
  \vspace{-1em}
\end{figure*}

\setcounter{table}{2}
\begin{table}[ht]
\centering
\caption{The ASR of AIA on all models in precisely recovering the seven PII types: fax, birthday, SSN, address, job, bitcoin, and UUID.}
\label{tab:pii_attack_result}
\resizebox{0.5\textwidth}{!}{
\begin{tabular}{cccccccc}
\toprule[2pt]
\multicolumn{1}{l}{}           & \textbf{fax}      & \textbf{birthday} & \textbf{SSN}      & \textbf{address}  & \textbf{job}      & \textbf{bitcoin}  & \textbf{UUID}     \\ \hline
\multicolumn{1}{c|}{GPT2-XL}   & 0.25     & 1.00     & 0.76     & 0.56     & 0.97     & 0.22     & 0.17     \\
\multicolumn{1}{c|}{Bloom-7B1} & 0.48     & 0.99     & 0.57     & 0.57     & 0.98     & 0.04     & 0.04     \\
\multicolumn{1}{c|}{LLaMA3-8B} & 0.20     & 0.95     & 0.38     & 0.41     & 0.89     & 0.03     & 0.10     \\ \bottomrule[1.5pt]
\end{tabular}
}
\vspace{-2em}
\end{table}

\noindent{\textbf{Results on various PII types.}}
Table~\ref{tab:pii_attack_result} presents the ASR of AIA in precisely recovering the seven PII types: fax, birthday, SSN, address, job, bitcoin, and UUID. Remarkably, the ASR for birthdays and jobs approaches 100\%. Birthdays, which are short and highly structured numerical sequences, likely benefit from the model's pre-training exposure to similar formats, resulting in minimal changes to their semantic encoding after fine-tuning. Jobs, typically consisting of one to three words, are relatively easier to recover compared to other PII types. This observation is further supported by the ROUGE-1 and BLEU-1 results on the PII dataset across different victim LLMs shown in Table~\ref{tab:base_result}.

All victim models exhibit strong recovery performance for PII types other than Bitcoin addresses and UUID, with recovery rates generally ranging from one-third to over half of the data. Owing to the inherent irregularity and extended length characteristics of Bitcoin addresses and UUIDs, precise reconstruction is significantly more challenging. Specifically, only the GPT2-XL model achieves a recovery rate of approximately 20\% for the two PII types, while the ASR for Bloom-7B1 and LLaMA3-8B remains below 10\%. Notably, even in cases of incomplete reconstruction, the generated outputs maintain substantial proximity to ground truth values, exhibiting only minor character-level discrepancies in alphanumeric sequences (e.g., single-letter substitutions or partial numeric mismatches).

Figure~\ref{fig:PII_attack_example} shows three comparison examples between the generated text and the original private data, with the quality of text reconstruction improving from left to right. The majority of common words and PII data can be precisely recovered, as indicated by the red highlights in the figure. However, the recovery of less frequent words (e.g., "Bitcoin") and special characters (e.g., "@") tends to be less successful. Additionally, the recovery of named entities may occasionally be imprecise. For long character sequences, such as phone numbers or UUIDs, over 80\% of the characters are typically recovered, although some minor errors in individual characters or capitalization issues may occur, as highlighted in purple in the figure.

\subsection{Ablation Study}
To explore the factors influencing the attack performance of AIA, we conducted three sets of ablation experiments on the decoder layer index, model size, and attack model architecture. The conclusions are as follows: 
\begin{itemize}[nolistsep, leftmargin=*, topsep=0pt]
    \item As the layer index increases, the attack performance decreases; however, the original private data can still be recovered to some extent. 
    \item The attack performance is independent of model size and AIA performs well in all model sizes.
    \item The attack performance is highly sensitive to the architecture of the attack model, with different architectures leading to poorer attack results.
\end{itemize}

\subsubsection{Decoder Layer Index}
\begin{figure}[t]
  \includegraphics[width=\linewidth]{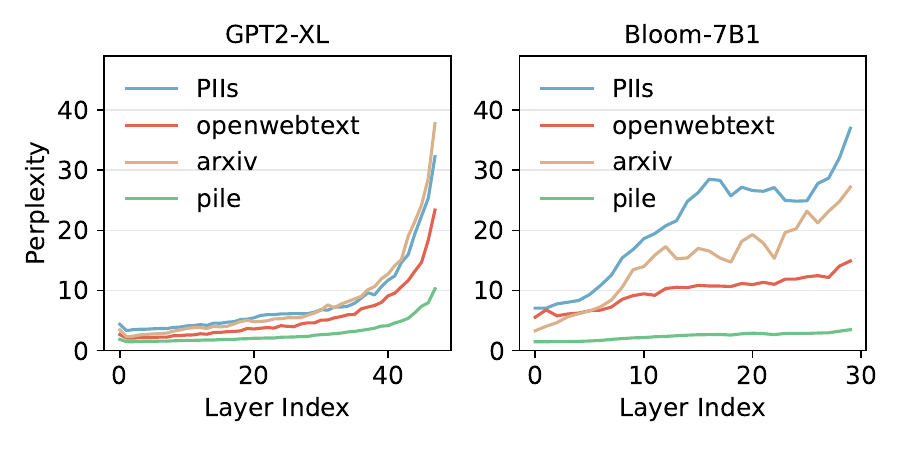} 
  \caption {The attack performance of AIA on GPT2-XL and Bloom-7B1 models as the attacker's decoder layer index varies, with the attack performance generally decreasing as the layer index increases.}
  \label{fig:layer_idx}
  \vspace{-1em}
\end{figure}

Figure~\ref{fig:layer_idx} illustrates the trend of PPL on GPT2-XL and Bloom-7B1 models as the attacker's decoder layer index varies. The results show that as the decoder layer index increases, i.e., as the data leakage layer moves closer to the output layers, the overall attack effectiveness declines. This observation aligns with the trend described in Section \ref{sec:act_cos}, where the cosine similarity of activations before and after fine-tuning decreases as the decoder layer index increases. The decline in attack performance can be attributed to the greater changes in the activations of the decoder layers that is closer to the output layer during fine-tuning. 

Interestingly, when the cosine similarity of activations before and after fine-tuning drops below 60\% for a particular decoder layer, the perplexity of the generated text remains below 40. This indicates that the generated sentences become less natural, with noticeable grammatical or contextual inconsistencies, which suggests a reduction in the fluency and coherence of the generated texts. However, despite these linguistic limitations, the attacker is still able to infer the original fine-tuning data to a certain extent. This highlights the robustness of AIA, even when the stage controlled by the attacker is positioned further back in the pipeline.


\subsubsection{Model Size}

\begin{table}[ht]
\centering
\caption{Attack performance of AIA on GPT-2 and Bloom models of varying sizes.}
\label{tab:model_size}
\resizebox{0.5\textwidth}{!}{
\begin{tabular}{ccccccc}
\toprule[2pt]
\multirow{2}{*}{\textbf{Victim Model}}                & \multirow{2}{*}{\textbf{Model Size}}                & \multirow{2}{*}{\textbf{Dataset}} & \multicolumn{4}{c}{\textbf{Metrics}}                  \\
                                             &                                            &                          & \textbf{PPL}      & \textbf{ROUGE-L}   & \textbf{BLEU-4}    & \textbf{COS}         \\ \hline
\multicolumn{1}{c|}{\multirow{12}{*}{Bloom}} & \multicolumn{1}{c|}{\multirow{4}{*}{560M}} & PIIs                     & 15.22   & 0.76    & 0.46    & 0.84     \\
\multicolumn{1}{c|}{}                        & \multicolumn{1}{c|}{}                      & openwebtext              & 4.06    & 0.94    & 0.75    & 0.92     \\
\multicolumn{1}{c|}{}                        & \multicolumn{1}{c|}{}                      & arxiv                    & 14.60   & 0.89    & 0.52    & 0.86     \\
\multicolumn{1}{c|}{}                        & \multicolumn{1}{c|}{}                      & pile                     & 2.46    & 0.97    & 0.88    & 0.94     \\ \cline{2-7} 
\multicolumn{1}{c|}{}                        & \multicolumn{1}{c|}{\multirow{4}{*}{1B7}}  & PIIs                     & 10.24   & 0.80    & 0.52    & 0.89     \\
\multicolumn{1}{c|}{}                        & \multicolumn{1}{c|}{}                      & openwebtext              & 3.31    & 0.96    & 0.81    & 0.95     \\
\multicolumn{1}{c|}{}                        & \multicolumn{1}{c|}{}                      & arxiv                    & 9.83    & 0.92    & 0.58    & 0.91     \\
\multicolumn{1}{c|}{}                        & \multicolumn{1}{c|}{}                      & pile                     & 2.01    & 0.98    & 0.92    & 0.96     \\ \cline{2-7} 
\multicolumn{1}{c|}{}                        & \multicolumn{1}{c|}{\multirow{4}{*}{7B1}}  & PIIs                     & 12.06   & 0.81    & 0.48    & 0.89     \\
\multicolumn{1}{c|}{}                        & \multicolumn{1}{c|}{}                      & openwebtext              & 4.41    & 0.96    & 0.81    & 0.95     \\
\multicolumn{1}{c|}{}                        & \multicolumn{1}{c|}{}                      & arxiv                    & 14.34   & 0.91    & 0.58    & 0.90     \\
\multicolumn{1}{c|}{}                        & \multicolumn{1}{c|}{}                      & pile                     & 1.92    & 0.98    & 0.91    & 0.96     \\ \hline
\multicolumn{1}{c|}{\multirow{12}{*}{GPT2}}  & \multicolumn{1}{c|}{\multirow{4}{*}{355M}} & PIIs                     & 5.70    & 0.80    & 0.52    & 0.75     \\
\multicolumn{1}{c|}{}                        & \multicolumn{1}{c|}{}                      & openwebtext              & 4.69    & 0.91    & 0.66    & 0.89     \\
\multicolumn{1}{c|}{}                        & \multicolumn{1}{c|}{}                      & arxiv                    & 11.84   & 0.87    & 0.54    & 0.86     \\
\multicolumn{1}{c|}{}                        & \multicolumn{1}{c|}{}                      & pile                     & 2.75    & 0.95    & 0.79    & 0.93     \\ \cline{2-7} 
\multicolumn{1}{c|}{}                        & \multicolumn{1}{c|}{\multirow{4}{*}{774M}} & PIIs                     & 4.30    & 0.81    & 0.56    & 0.82     \\
\multicolumn{1}{c|}{}                        & \multicolumn{1}{c|}{}                      & openwebtext              & 3.42    & 0.93    & 0.71    & 0.91     \\
\multicolumn{1}{c|}{}                        & \multicolumn{1}{c|}{}                      & arxiv                    & 8.79    & 0.90    & 0.60    & 0.88     \\
\multicolumn{1}{c|}{}                        & \multicolumn{1}{c|}{}                      & pile                     & 2.30    & 0.96    & 0.84    & 0.94     \\ \cline{2-7} 
\multicolumn{1}{c|}{}                        & \multicolumn{1}{c|}{\multirow{4}{*}{1.5B}} & PIIs                     & 3.44    & 0.85    & 0.62    & 0.90     \\
\multicolumn{1}{c|}{}                        & \multicolumn{1}{c|}{}                      & openwebtext              & 3.62    & 0.95    & 0.76    & 0.94     \\
\multicolumn{1}{c|}{}                        & \multicolumn{1}{c|}{}                      & arxiv                    & 5.16    & 0.92    & 0.67    & 0.92     \\
\multicolumn{1}{c|}{}                        & \multicolumn{1}{c|}{}                      & pile                     & 1.65    & 0.97    & 0.89    & 0.96     \\ \bottomrule[1.5pt]
\end{tabular}
}
\vspace{-1em}
\end{table}
Table~\ref{tab:model_size} systematically presents the experimental results for GPT2 and Bloom models with varying parameter scales. To ensure comprehensive experiments, we select three representative configurations for each model family: the GPT2 series includes 355M, 774M, and 1.5B parameter variants, while the Bloom series comprises 560M, 1.7B, and 7.1B parameter configurations. 
The experimental results demonstrate that the attack performance of AIA is highly dependent on the victim dataset, and it maintains stable performance across different model sizes, with most PPL consistently below 10, ROUGE-L scores exceeding 0.9, and BLEU-4 scores above 0.6 in most cases. 


\subsubsection{Attack Model Architecture}
To explore the impact of the attack model architecture on attack performance, we conduct experiments using Mistral~\cite{mistral} and Qwen2.5~\cite{qwen2.5} as attack model architectures and compare them to the victim model architecture. Each attack model is configured with six decoder layers. As shown in Table \ref{tab:attack_model}, while all attack models exhibit excellent performance when trained on the shadow dataset, their effectiveness significantly declines when transitioning to inverting the victim dataset after switching the attack model architecture. Notably, even the best-performing configuration on GPT2-XL still yields perplexity values ranging from 24 to 120. On the Bloom-7B1 and LLaMA3-8B models, the perplexity can even reach values above a thousand, rendering AIA almost completely ineffective.



\section{Conclusion}
In this paper, we explore the privacy risks inherent in decentralized training, particularly in scenarios where an honest-but-curious attacker exists in the pipeline. Despite lacking access to the complete model weights, we demonstrate the feasibility of simulating the victim model using a pre-trained model and introduce Activation Inversion Attack (AIA). We conduct extensive experiments on various large language models and public datasets to emphasize the effectiveness of our attack. As the application of decentralized training continues to grow, we call for the development of effective defense measures to mitigate the risk of AIA.

\section*{Limitations}
Our method has a key limitation: the architecture of the attack model must be consistent with that of the clean model. While the attack model performs well on the shadow dataset when using different architectures, its effectiveness significantly decreases when applied to the clean dataset. This constraint limits the flexibility in choosing the attack model. Additionally, the generated text exhibits issues such as lack of fluency, inconsistencies in letter casing, errors with special characters, uncommon words, and difficulty in accurately recovering long sequences. These observations indicate that our method is influenced by the challenges of transferring to unknown data distributions and the variations introduced during model fine-tuning.

\section*{Ethics Statement}
We declare that all authors of this paper adhere to the ACM Code of Ethics and uphold its code of conduct. This paper investigates activation inversion attack in decentralized training. The objective of our work is to highlight the potential data leakage risks associated with decentralized training, aiming to encourage the community to give greater attention to privacy protection in such settings and to advocate for measures to prevent such information leaks. No real sensitive data is used in our experiments; all experiments are conducted with publicly available datasets. The data in the PII dataset we use is randomly generated and does not represent actual private information. All models employed in this study are open-source and thus do not pose any threat to proprietary models.



\newpage
\appendix

\section{Hyperparameters}
During the training of the attack model, the sequence length is set to 160. For fine-tuning the victim models, the sequence length is set to 1600 for LLaMA3-8B and Bloom-7B1, and 800 for GPT2-XL. The AdamW\cite{loshchilov2017decoupled} optimizer is used for all training and fine-tuning processes, with learning rates set to 5e-5 for GPT2-XL and Bloom-7B1, and 7e-5 for LLaMA3-8B, along with an epsilon value of 1e-8.

\section{Datasets}
The WikiText dataset serves as a high-quality, clean, and large-scale collection of English text extracted from Wikipedia articles, providing a solid foundation for creating the shadow dataset for the attacker's model. The ArXiv dataset is a large-scale collection of scientific papers from the arXiv repository. The OpenWebText dataset is a high-quality, large-scale corpus of English web content, curated from URLs shared on Reddit with high karma. The Pile is an 800GB, diverse English text dataset designed for training large language models, combining content from 22 high-quality sources, including books, academic papers, code, and web text. The PII dataset consists of 1,000 instances of sensitive information and includes 10 types of personally identifiable information (PII), such as phone numbers, email addresses, and home addresses, presented in a structured format. These data are randomly generated using regular expressions and do not represent real private information.

\section{Toolkits}
We use the NLTK package to measure the BLEU score, the rouge\_score library to calculate the ROUGE score, and scikit-learn to compute the cosine similarity.

\section{True-Prefix and SPT Attack Examples}
\begin{figure}[H]
  \includegraphics[width=\linewidth]{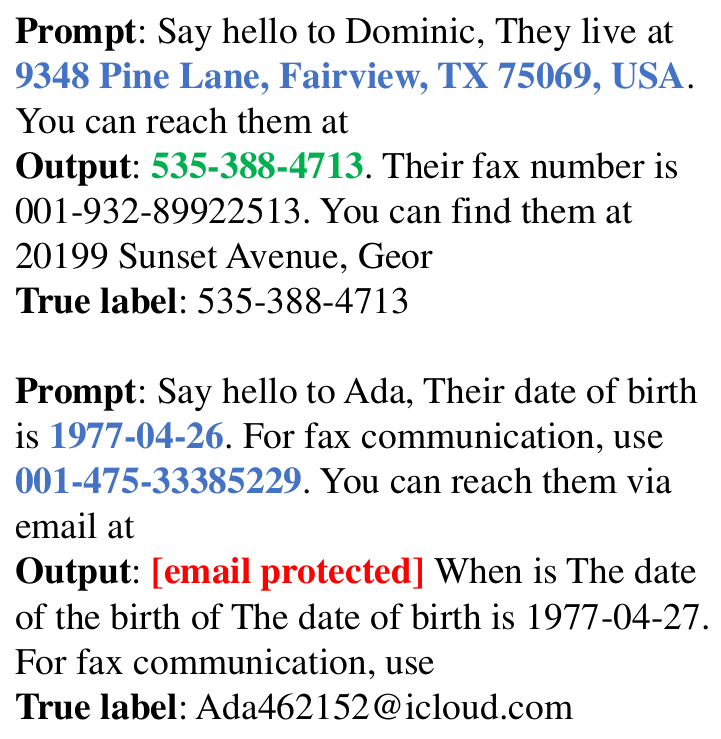} 
  \caption {Two True-Prefix attack examples. Blue text represents the real private data, while green and red text indicate successful and failed privacy theft, respectively.}
  \label{fig:true_prefix_example}
  \vspace{-1em}
\end{figure}

\begin{figure}[H]
  \includegraphics[width=\linewidth]{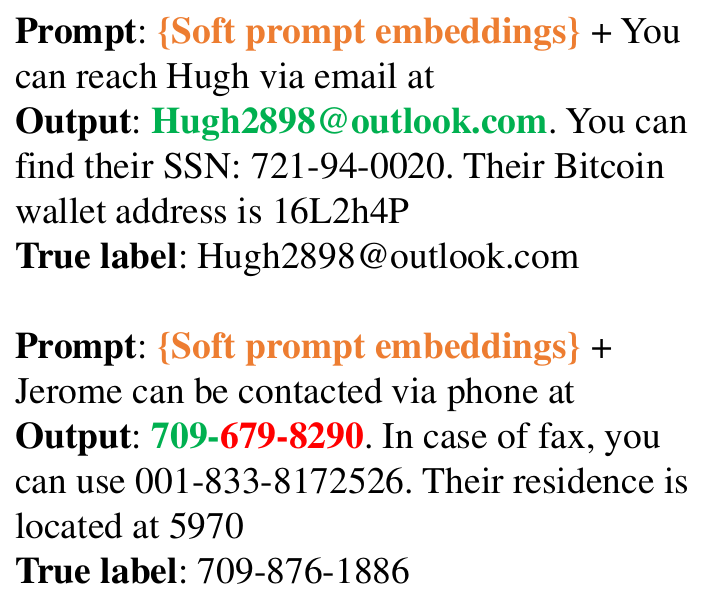} 
  \caption {Two SPT attack examples. Orange text represents the soft prompt embeddings, with green and red text indicating successful and failed privacy theft, respectively.}
  \label{fig:spt_example}
  \vspace{-1em}
\end{figure}

Figure \ref{fig:true_prefix_example} and Figure \ref{fig:spt_example} present two examples of True-Prefix~\cite{true-prefix} and SPT~\cite{SPT} attacks, respectively. In the True-Prefix attack, we insert real data of additional PII types, such as address or birthday, before the prompt templates, as shown in the blue sections in Figure \ref{fig:true_prefix_example}. In the SPT attack, we train on 64 PII data pairs for 5 epochs to obtain the soft prompt embeddings, which are set to a length of 10. The soft prompt embeddings are then concatenated before the prompt templates. During the training, the victim model remains frozen, with no gradient updates applied.

\end{document}